\documentclass[12pt]{article}
\usepackage{latexsym}
\usepackage{graphicx}

\thicklines                         
\long \def \blockcomment #1\endcomment{}

\begin{document}           

\baselineskip=0.33333in
\begin{quote} \raggedleft TAUP 2846-2007
\end{quote}
\vglue 0.5in

\begin{center}{\bf Experimental Consequences of \\
Regular Charge-Monopole Electrodynamics}
\end{center}
\begin{center}E. Comay$^*$
\end{center}

\begin{center}
School of Physics and Astronomy \\
Raymond and Beverly Sackler Faculty of Exact Sciences \\
Tel Aviv University \\
Tel Aviv 69978 \\
Israel
\end{center}
\vglue 0.5in
\vglue 0.5in
\noindent
\vglue 0.2in
\noindent
Abstract:

This work points out an application of the regular
theory of charge-monopole
electrodynamics to the new experiments of very high energy. Using this theory,
it is explained why the new regions of high energy
interactions that will be explored by CERN's
Large Hadron Collider may show a significant increase of the
cross section. This conclusion illustrates the significance of the
extension of electrodynamics that takes the form of the regular
charge-monopole theory.

\newpage
\noindent
{\bf 1. Introduction}
\vglue 0.33333in

In physics, theoretical predictions of new experimental results
rely mainly on the
mathematical structure of the relevant theory. In this sense,
the activity in theoretical physics is analogous to that of mathematics.
However, unlike problems belonging to
the realm of mathematical investigation, there are
physical problems where
the theoretical analysis cannot proceed without
a knowledge of some experimental data that
describe specific properties of the system.
These two elements are used below in an explanation of the possibility that
the data which will be obtained from CERN's Large Hadron Collider (LHC)
may show a significant increase of the cross section for regions of
very high energy that have not been explored before.

\vglue 0.66666in
\noindent
{\bf 2. The Experimental Data}
\vglue 0.33333in

Preliminary results of the experimental data needed
for the analysis have been reported nearly 10 years ago [1,2].
These experiments have been carried out at HERA and use $e^+p$
colliding beams. Both H1 and ZEUS
experiments show a significant excess of the number of events
for regions where $Q^2 > T$ GeV$^2$. Here
$Q^2$ is the absolute value of the squared 4-momentum transferred in
the interaction and $T$ denotes the threshold used, which is
15000, 35000 GeV$^2$, respectively. The validity of this
result is crucial for the discussion carried out
below. The main problem with the data reported in [1,2] is
the small number of events found in the analysis. Thus, different
analyses use different cuts of the kinematic region and the number
of events found ranges between 2 and 12 [1,2]. Hence, more data
are needed.

Further publications that use a larger set of data obtained
from the HERA facility have been
published since then. It turns out that recent publications [3,4]
which rely on all
the relevant HERA data still leave this issue undecided. Thus,
unlike the first publications that use $Q^2$ as a parameter for
grouping the data, the
new analyses use the value of the transverse momentum $p_T$.
(In the latter calculations the contribution of the longitudinal
component of $Q^2$ is lost and it becomes more difficult to
detect events restricted to very high $Q^2$.)
The recent analyses [3,4] show that the two HERA
groups, H1 and ZEUS, do not agree now on the issue described above and
that the excess of the number of events found is still too small.

This state of affairs explains why today it is not clear whether or
not there is a significant increase of the cross section for
events where $Q^2$ is very high.

The foregoing discussion explains why the following statement takes
the status of a conjecture.

\begin{itemize}
\item[{A.}] The increase of the cross section reported in [1,2]
is just the tip of the iceberg. A significant increase will be found in
the data of more energetic interactions.
\end{itemize}

Henceforth, this statement is called conjecture A. It is explained
why also the theoretical analysis carried out below indicates
that several alternatives are possible and that conjecture A is
consistent with one of them.

The HERA facility contains a beam of protons that collide with a beam
of leptons. The leptonic beam is made of electrons or positrons. In
the theoretical analysis described below it is explained why results of the
proton-proton LHC facility
are relevant to the problem presented above.

\vglue 0.66666in
\noindent
{\bf 3. A Theoretical Discussion}
\vglue 0.33333in

The theory used herein is the Regular Charge-Monopole Theory (RCMT)
[5,6] and its application to strong interactions [7].
Let us begin with a brief description of relevant properties of RCMT.
As usual, the term ``monopole" is used sometimes instead of
``magnetic charge".
Thus, monopoles are defined by the following duality transformation (called
also duality rotation by $\pi /2$)
\begin{equation}
\bf E \rightarrow \bf B,\;\;\;\bf B \rightarrow -\bf E
\label{eq:EB}
\end{equation}
and
\begin{equation}
e \rightarrow g,\;\;\;g\rightarrow -e,
\label{eq:EG}
\end{equation}
where $g$ denotes the magnetic charge of monopoles.

The structure of RCMT [5,6] conserves
the duality relations between two electromagnetic theories:
the theory of electromagnetic fields
and a matter carrying electric charge (and no magnetic charge) -
namely, the ordinary Maxwellian electrodynamics and the corresponding theory
of fields and a matter carrying magnetic charge
(and no electric charge). Duality conservation is consistent with
a self-evident requirement
imposed on a monopole theory. RCMT is the union of these subtheories which
conserves the form of the charge-monopole duality relations.
This theory can be derived from a regular Lagrangian density. Hence,
the corresponding quantum mechanical theory can be constructed in a
straightforward manner. The main result of RCMT
can be put in the following words:
Charges do not interact with bound fields of monopoles and
monopoles do not interact with bound fields of charges. Charges interact
with all fields of charges and with radiation fields emitted from
monopoles. Monopoles interact with all fields of monopoles and with
radiation fields emitted from charges. Another important result of
RCMT is that the unit of the elementary magnetic charge $g$ is a
free parameter. However, hadronic data indicate that
this unit is much larger than that of the electric charge.
More details of RCMT can be found in [5-7].

These properties of RCMT
fit like a glove the data of electromagnetic projectiles
interacting with nucleons (see [7], pp. 90, 91).
Thus, protons and neutrons do not look alike
in cases of charged lepton scattering whereas they look very similar
if the projectile is a hard enough photon [8]. It is also shown [7,9,10] that
an application of RCMT provides explanation for other physical properties
in general and for hadronic properties in particular. Thus,
explanations are obtained for the following properties:
the reason why magnetic monopoles have not been detected in our instruments;
the uniform density of nuclear matter; the fact that a system of six
valence quarks is a deuteron, which is a
proton-neutron loosely bound state;
the fact that {\em strongly} bound states of pentaquarks do not exist;
the reason why antiquarks of nucleons occupy a larger spatial volume
than that of quarks;
the EMC effect [11,12] which shows that in a nucleus, the volume of a
single quark increases together
with the increase of the nucleon number A.

These results encourage one to continue the examination of the relevance
of RCMT to the baryonic structure. The main baryonic property inferred
from RCMT is that baryons have a core whose magnetic charge is $3g$
and each quark has one negative unit of magnetic charge $-g$. Therefore,
baryons are neutral with respect to magnetic charge. (The
existence of a baryonic core is consistent with the portion of
momentum of a very energetic nucleon, which is carried by
quarks [13].) At this
point, RCMT can say nothing more on the structure of the baryonic core.
In particular, it is not clear whether the core is an
elementary structureless object
or a complicated system which contains closed shells of quarks.
However, the following arguments support the second alternative:

\begin{itemize}
\item An elementary structureless baryonic core is certainly
a simpler object than a core which contains closed shells of quarks.
Thus, if one does not expect to find that Nature is too simple then
he should be ready to realize that the baryonic core has closed shells
of quarks.

\item Here a comparison of atomic size with that of the nucleon and of the
$\pi $ meson provides another hint. The size of the radial variable of
the positronium's ground state
is very nearly two times larger than that of the hydrogen atom ground state.
However, for the hydrogen atom the classical meaning of
this variable represents the Bohr atomic radius $r=0.53$ \AA.
On the other hand, in
the case of the positronium it represents the diameter
of the system. Thus, one
concludes that the geometrical
size of the positronium's ground state is roughly the same as that of the
hydrogen atom.

Now let us examine the helium atom. Here the strength of the
Coulomb attraction of the nucleus is twice as large as that of the
hydrogen atom and the Pauli exclusion principle allows the two electrons
to be in the lowest orbital. Thus, due to the stronger attraction of
the nucleus, one finds that the radius of the ground state of the
helium atom, $r=0.31$ \AA,
is considerably smaller than that of the corresponding state of the
hydrogen atom and of the positronium as well.

Let us turn to the corresponding hadronic data and examine the
$\pi $ meson and the proton. Here the $\pi $ meson, which
consists of a $\bar {q}q$ pair, corresponds to
the positronium's ground state. In the case of the proton,
we have $uud$ quarks and the Pauli exclusion principle allows all
these quarks to be in the lowest empty orbital. Thus, if the analogy with
the atomic data mentioned above is relevant to this case then one
expects that the size of the proton should be considerably smaller than
that of the $\pi $ meson. The experimental data is inconsistent with
this expectation and the effective radius of the $\pi $ meson is
smaller than that of the proton
$r_\pi \simeq 0.8\,r_p $.

These experimental data can be interpreted as follows. The proton (as well
as all other baryons) has a core charged with monopole units {\em and}
inner closed shells of quarks. The net monopole units of the core
and the closed shells of quarks is +3 and it attracts the three valence
quarks. The entire system is neutral with respect to magnetic monopole
charge. Since the orbitals
of the three valence quarks are orthogonal to those of the closed
shells, the former are pushed outwards.
This effect explains the rather large geometrical size of the proton.
\end{itemize}

Following these arguments, it is assumed here that
\begin{itemize}
\item[{B.}] Baryons have a core which contains closed shells of quarks.
\end{itemize}
Clearly, the previous statement is a conjecture which is called below
conjecture B. The following discussion shows how conjecture B can
explain conjecture A.

As of today, the energies used in collisions which involve nucleons are
analyzed by the interaction of
the valence quarks and the associated $\bar {q}q$ pairs [13].
Now, according to conjecture B, the closed shells of quarks at the
baryonic core make a quantum mechanical bound system. Thus, in cases
of interactions with projectiles having a relatively low energy,
the baryonic core behaves like an inert object. On
the other hand, interactions that involve a high enough
energy may excite quarks of the baryonic core. This scenario is analogous
to the effect demonstrated by the Franck-Hertz experiment
which is known for more than 90 years [14]. Thus, if the
projectile has enough energy then quarks of the outer closed shell of the
baryonic core begin to participate in the scattering process. The
contribution of the new participants should increase the number of events.

Now, interactions of the projectiles with the three valence quarks
(and with the additional $\bar {q}q$) explain the
data of the lower $Q^2$ part of the HERA experiments
whereas an excess of events is found for the higher $Q^2$
values [1,2]. According to conjecture A, this effect is significant.
Thus, these reports
indicate that the energies of the HERA facility
are just above the threshold of the
energy required for exciting quarks of the
closed baryonic shells. Evidently, due to phase space consideration,
one expects that the excess of the number of events will be more
significant for energies which are higher than those of the HERA
facility.

\vglue 0.66666in
\noindent
{\bf 4. Concluding Remarks}
\vglue 0.33333in

The theoretical analysis of the previous Section examines the conditions
required for the excitation of the closed shells of quarks
belonging to the baryonic core. Therefore,
it is relevant to scattering processes that include baryons.
For this reason, the data obtained from the electron-positron
CERN's LEP collider cannot cast light on this problem. Similarly, the
TEVATRON proton-antiproton collider cannot be used for this purpose.
Indeed, by definition, in a particle-antiparticle collision, {\em all}
constituents participate in the mutual annihilation.
This property is
independent of the collision energy. It follows that
the quark analog of the Franck-Hertz effect cannot be seen in this case.
Restricting the discussion to facilities that have been used till now,
one finds that the
HERA facility is the sole source of data required for a confirmation
of conjecture B of Section 3.
The HERA data provide us with no more than a hint
indicating that for higher energies, the excess of the number of
events may be significant [1,2]. This hint enables one to make
conjecture A of Section 2.

The operation of the
LHC will yield data of proton-proton collisions at very high energy.
This experiment differs from those of HERA, where a proton collides with
an electron (positron). However, if conjectures A, B hold then
the analogue of the Franck-Hertz
effect of the closed shells of quarks should be seen in the LHC data.

At the time when this work is finished the LHC has not yet begun to work.
Hence, at present, it is still not known whether its data will support
conjecture A.
As explained in the previous Section, an experimental verification
of conjecture A supports conjecture B stating that
the baryonic core contains closed shells of quarks. Three
different kinds of explanation can be used for a case where the LHC
data will not support this prediction: the regular charge-monopole theory
is irrelevant to baryons; the baryonic core is an elementary structureless
object which contains no closed shells of quarks;
closed shells of the baryonic core exist but the LHC
energy is lower than the threshold required for their excitation. Let
us wait and see.


\newpage
References:
\begin{itemize}
\item[{*}] Email: elic@tauphy.tau.ac.il  \\
\hspace{0.5cm}
           Internet site: http://www-nuclear.tau.ac.il/$\sim $elic

\item[{[1]}] C. Adloff et al., Z. Phys {\bf C74}, 191 (1997).
\item[{[2]}] J. Breitweg et al., Z. Phys {\bf C74}, 207 (1997).
\item[{[3]}] D. M. South, hep-ex/0605119.
\item[{[4]}] C. Diaconu, hep-ex/0610041.
\item[{[5]}] E. Comay, Nuovo Cimento, {\bf 80B}, 159 (1984).
\item[{[6]}] E. Comay, Nuovo Cimento, {\bf 110B}, 1347 (1995).
\item[{[7]}] E. Comay,
in {\em Has the Last Word Been Said on Classical Electrodynamics?} Edited
by A. Chubykalo, V. Onoochin, A. Espinoza, and R. Smirnov-Rueda (Rinton
Press, Paramus, NJ, 2004). (The article's title is
``A Regular Theory of Magnetic Monopoles and
Its Implications.")
\item[{[8]}] T. H. Bauer, R. D. Spital, D. R. Yennie and
F. M. Pipkin, Rev. Mod. Phys. {\bf 50}, 261 (1978).
\item[{[9]}] E. Comay, Apeiron {\bf 13}, 162 (2006).
\item[{[10]}] E. Comay, Lett. Nuovo Cimento, {\bf 43}, 150 (1985).
\item[{[11]}] J. J. Aubert et al. (EMC), Phys. Lett., {\bf 123B}, 275 (1983).
\item[{[12]}] A. Bodek et al., Phys. Rev. Lett., {\bf 50}, 1431 (1983).
\item[{[13]}] D. H. Perkins, {\em Introductions to High Energy Physics}
3rd edn. (Addison-Wesley, Menlo Park, 1987). (See pp. 275-283).
\item[{[14]}] A. Messiah, {\em Quantum Mechanics} (North Holland,
Amsterdam, 1961). p. 24.

\end{itemize}

\end{document}